\documentclass[twocolumn,prb,amsmath,showpacs,amssymb,superscriptaddress,a4paper]{revtex4}

\usepackage{graphicx}
\usepackage{dcolumn}
\usepackage{bm}
\usepackage[sort&compress]{natbib}

\newcommand{\grad}{\ensuremath{^\circ}}
\newcommand*{\unit}[1]{\,\mathrm{#1}}

\newcommand{\coob}[1]{$\mathbf{CoO_{2}}$~}
\newcommand{\coo}[1]{$\mathrm{CoO_{2}}$~}
\newcommand{\naxb}[1]{$\mathbf{Na_{x}CoO_{2}}$~}
\newcommand{\nax}[1]{$\mathrm{Na_{x}CoO_{2}}$~}
\newcommand{\naxs}[1]{$\mathrm{Na_{x}CoO_{2}}$}
\newcommand{\naab}[1]{$\mathbf{Na_{0.8}CoO_{2}}$~}
\newcommand{\naa}[1]{$\mathrm{Na_{0.8}CoO_{2}}$~}
\newcommand{\naas}[1]{$\mathrm{Na_{0.8}CoO_{2}}$}
\newcommand{\nabb}[1]{$\mathbf{Na_{0.85}CoO_{2}}$~}
\newcommand{\nabbs}[1]{$\mathbf{Na_{0.85}CoO_{2}}$}
\newcommand{\nab}[1]{$\mathrm{Na_{0.85}CoO_{2}}$~}
\newcommand{\nabs}[1]{$\mathrm{Na_{0.85}CoO_{2}}$}
\newcommand{\nac}[1]{$\mathrm{Na_{0.3}CoO_{2}}$~}
\newcommand{\nad}[1]{$\mathrm{Na_{0.7}CoO_{2}}$~}
\newcommand{\naes}[1]{$\mathrm{Na_{0.5}CoO_{2}}$}

\begin{document}

\title{Spin fluctuations, magnetic long-range order and Fermi surface gapping in $\mathbf{Na_{x}CoO_{2}}$}

\author{T.~F.~Schulze}
 \thanks{corresponding author, email: tschulze@phys.ethz.ch, present address: Helmholtz-Zentrum Berlin f\"ur Materialien und Energie, department SE1, Kekul$\mathrm{\acute e}$stra\ss e 5, D-12489 Berlin, Germany}
\author{M.~Br\"uhwiler}
\author{P.~S.~H\"afliger}
\author{S.~M.~Kazakov}
 \thanks{present address: Chemistry Department, Moscow State University, 119991, Moscow, Russia}
\affiliation{Laboratory for Solid State Physics, ETH Z\"urich, CH-8093 Z\"urich, Switzerland}
\author{Ch.~Niedermayer}
 \affiliation{Laboratory for Neutron Scattering, ETH Z\"urich \& Paul Scherrer Institut (PSI), CH-5232 Villigen, Switzerland}
\author{K.~Mattenberger}
\author{J.~Karpinski}
\author{B.~Batlogg}
\affiliation{Laboratory for Solid State Physics, ETH Z\"urich, CH-8093 Z\"urich, Switzerland}

\date{\today}

\begin{abstract}
In this study an extended low energy phase diagram for \nax\ is experimentally established with emphasis on the high $x$ range. It is based on systematic heat capacity studies on both polycrystalline and single crystalline samples and on $\mathrm{\mu SR}$ measurements. Main features are the existence of mass enhancement, spin fluctuations without long-range order, and magnetic order with associated Fermi surface gapping. The latter is seen in the electronic density of states (DOS) and suppression of nuclear specific heat. While there is agreement between the band structure and the low energy DOS in the low $x$ range, in the high $x$ range ($x \geq 0.6$) the thermodynamically determined DOS is approximately three times that deduced from the angle-resolved photoemission spectroscopy (ARPES)-measured band dispersion or local-density approximation (LDA) calculations.
\end{abstract}


\maketitle

\section{Introduction}
The layered transition metal oxide $\mathrm{Na_{x}CoO_{2}}$ combines high thermopower and metallic conductivity, making it a promising candidate for thermoelectric energy conversion \cite{ter97,lee06}. The discovery of superconductivity in hydrated compounds \cite{tak03} and other types of ordered electronic groundstates for higher sodium content \cite{foo04,boo04,sal04,bay04,woo05} have intensified the interest and raised questions related to Fermi surface (FS) instabilities. Itinerant electrons moving on the $\mathrm{CoO_{2}}$ triangular lattice and the ability to control the filling of the narrow Co-O derived bands through the sodium content $x$ render it an appealing model system in the field of correlated electron physics.

It is especially interesting to address the low energy electronic excitation density as it is both a measure of correlation effects and an indicator of collective groundstates. In order to explore the magnetic and thermodynamic properties of the electronic system in this context in detail, we studied the heat capacity of a large number of both polycrystalline and single crystalline samples and performed Muon Spin Rotation ($\mathrm{\mu SR}$) measurements. Our data, combined with previous results, provide a clear picture across the whole sodium concentration range: In addition to the charge-ordered insulator at $x=0.5$ we observe a strong enhancement of the linear specific heat coefficient $\gamma$ for $x>0.6$ which might be taken as evidence for either further enhanced correlation effects or for additional bands crossing $E_{F}$. Collective magnetic groundstates of spin-density wave (SDW) type with various ordering temperatures are observed in the high $x$ range accompanied by pronounced Fermi surface gapping. Additional evidence for the loss of DOS stems from the low-temperature (low-$T$) nuclear heat capacity. The magnetic instabilites are foreshadowed by spin fluctuations without magnetic long-range order at slightly lower $x$, which manifest themselves in an even more strongly enhanced and field-dependent $\gamma$ at lowest temperatures ($T<10\unit{K}$). 

\section{Experimental Details and Methods}
Polycrystals were prepared employing the conventional synthesis method involving either slow heating or the 'rapid heat-up'
technique \cite{mot01}. A stoichiometric mixture of high-purity $\mathrm{Co_{3}O_{4}}$ nanopowder (Aldrich, $99.995\%$) and waterfree $\mathrm{Na_{2}CO_{3}}$ (Aldrich, $99.995\%$) was pressed into pellets and annealed overnight at $750...850\unit{\grad C}$ in air. The samples were confirmed to be a single phase of hexagonal \nax\ by X-ray powder diffraction. The Na content of the polycrystals was determined from the unit-cell parameters, in particular by the $c/a$ ratio. As the $c/a$ vs. $x$ curve becomes flat for $x>0.85$, this method of sodium content determination is inaccurate in this range. Thus, $x$ values higher than $x=0.85$ have to be considered with caution and are in fact most likely not representative of the true Na content. This will be taken into account in the interpretation of the data.

Single crystals were grown applying a standard floating zone technique:
First, \nax\ powder was prepared by direct reaction of $\mathrm{Co_{3}O_{4}}$ and $\mathrm{Na_{2}CO_{3}}$ (the same starting materials as mentioned above) in an alumina boat under oxygen flow at $650\unit{\grad C}$. Rapid heating of the furnace ($2\unit{\grad C/min}$) to the desired sintering temperature reduced Na evaporation. After $4...6\unit{h}$ of sintering the resulting polycrystalline samples were ground under argon atmosphere and heated again for $16\unit{h}$ under oxygen flow at $975\unit{\grad C}$. The feeding material for single crystal growth were pressed pellets sintered at $975\unit{\grad C}$ for $8\unit{h}$. Using a four-mirror optical floating-zone furnace (Crystal Systems Corp.) we grew single crystals under flow of a mixture of argon and oxygen at $5\unit{atm}$. A more detailed preparation procedure can be found in the literature \cite{prab04}. Single crystals were analyzed using single crystal X-ray diffractometry and von-Laue backscattering which generally confirmed the samples to be single-crystalline, the mosaic spread being $\leq1.5\grad$. Some samples showed twinning, but in all cases the c-axis and the ab-planes were well-defined. The Na content of the crystal batches was determined by inductively coupled plasma atomic emission spectroscopy (ICP-AES). The study comprised more than 30 samples from different batches.

The specific heat was measured in a Quantum Design PPMS-14 and some measurements made use of the PPMS He-3 option. Addenda measurements and field-dependent thermometer calibration ensured reliability of the $C_{p}$ measurements.
$\mathrm{\mu SR}$ measurements were performed at the $\pi M3$ beamline at the Paul-Scherrer-Institut (PSI) at Villigen, Switzerland. This beamline provides $100\unit{\%}$ polarized muons.

In the following we contrast the linear specific heat coefficient $\gamma$, which is a measure for the low energy electronic excitation density, with $\gamma$ as calculated from results of numerous previous angle-resolved photoemission spectroscopy (ARPES) studies. Only in a few reports the value for the density of states $\gamma$ is given explicitly, but it can readily be determined from the Fermi velocity and wave vector assuming a 2D Fermi surface topology. Deviations from an exact 2D nature of the electronic structure have been discussed for higher $x$ as e.g.~a 3D magnetic structure was observed \cite{bay05,hel06} or tight-binding calculations yielded a significant $z$-dispersion \cite{joh04}. Other theoretical studies challenge these findings based on the assumption of strong correlations or an AFM groundstate \cite{geck07}. To calculate a $\gamma$ value from the Fermi surface observed in ARPES we follow the approach of a quasi-2D electronic structure which is straightforward because of the quasi-2D FS topology observed in almost all ARPES studies and which has previously been established to be applicable for \naxs\ \cite{qui06c}:
\begin{equation}
\gamma=\frac{\pi N_{A} k_{B}^{2} a_{0}^{2}}{3\hbar^{2}}\sum m^{\ast} \quad \textrm{where} \quad m^{\ast}=\frac{\hbar k_{F}}{v_{F}},
\label{eq:gamma1}
\end{equation}
Thus we can calculate $\gamma$ corresponding to the measured Fermi velocity $v_{F}$ and the Fermi wave vector $k_{F}$. In some measurements \cite{yang04,geck07} an anisotropy of the Fermi velocity by factor 2 was noted. We take this into account by stating the maximum and minimum DOS that would result from both extremes in these cases (those values are linked by broken lines in the phase diagram, Fig.~\ref{fig:phasediag}). As will be discussed later, $\gamma$ as obtained from ARPES is significantly lower than the thermodynamically determined $\gamma$ which can be interpreted as evidence for the existence of FS parts and/or a pronounced three-dimensionality of the FS not seen in ARPES.

\begin{figure}
\includegraphics[width=0.45\textwidth]{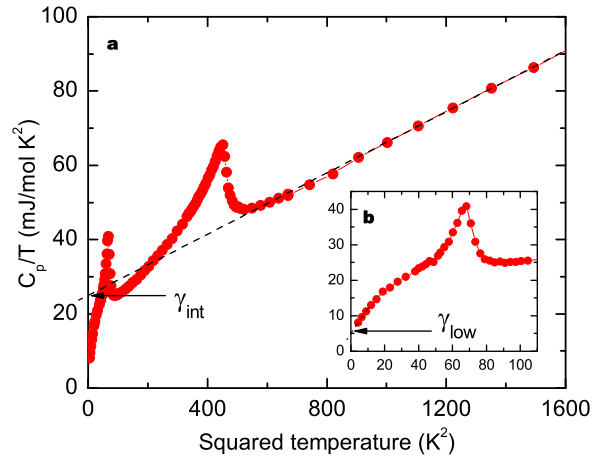}
\caption{\label{fig:cphighlow} Typical specific heat data for a slowly cooled \nab\ single crystal with two transitions into a long-range ordered magnetic state (at $22\unit{K}$ and $8\unit{K}$). (a) The electronic density of states at intermediate temperatures $\gamma_{int}$ (above any Fermi surface instabilities) is determined by a linear fit to the $C_{p}/T$ vs.~$T^{2}$ data. The residual low-$T$ DOS ($\gamma_{low}$) is determined from a linear extrapolation of the low-$T$ data down to $T \rightarrow 0K$ (panel (b)). }
\end{figure}

In this study we distinguish between the 'intermediate temperature DOS' $\gamma_{int}$ and the 'low temperature DOS' $\gamma_{low}$, both measured via the specific heat. The first characterizes the DOS at T above any magnetic order or spin fluctuation effects ($T>25\unit{K}$) and is obtained either through a fit of the $C_{p}$ data using the standard Debye law $C_{p}=\gamma T + \beta T^{3}$ for $T=30...60\unit{K}$ or a model comprising an Einstein and a Debye contribution
\begin{eqnarray}
C_{p} = \gamma T + 9 N_{D} k_{B} \left(\frac{T}{\theta_{D}}\right)^3 \int^{\theta/T}_{0}\frac{x^4 e^x}{\left(e^x-1\right)^2} \; dx \nonumber \\
+ \; 3 N_{D} k_{B} \left(\frac{\theta_{E}}{T}\right)^2 \frac{e^{(\theta_{E}/T)}}{\left(e^{(\theta_{E}/T)}-1\right)^2}
\label{eq:debye2}
\end{eqnarray}
to take into account the two prominent contributions to the phonon spectrum for $T=30...300\unit{K}$ \cite{lynn03}. We have analyzed the data of several samples using the two models and found excellent agreement with both. The resulting value for $\gamma_{int}$ is essentially unaffected by the choice of the phonon model. Thus, the Debye law fit is sufficient for the present purpose and it was therefore generally applied.

$\gamma_{low}$ is a measure for the excitation density in a $T$ range where magnetic order and spin fluctuations contribute to $C_{p}$, and $\gamma_{low}$ is obtained through extrapolation of the low-$T$ $C_{p}$ data down to $T \rightarrow 0K$. Fig.~\ref{fig:cphighlow} shows an example of both extrapolation procedures for a \nab\ single crystal (low temperature details in the inset).

\begin{figure}
\includegraphics[width=0.45\textwidth]{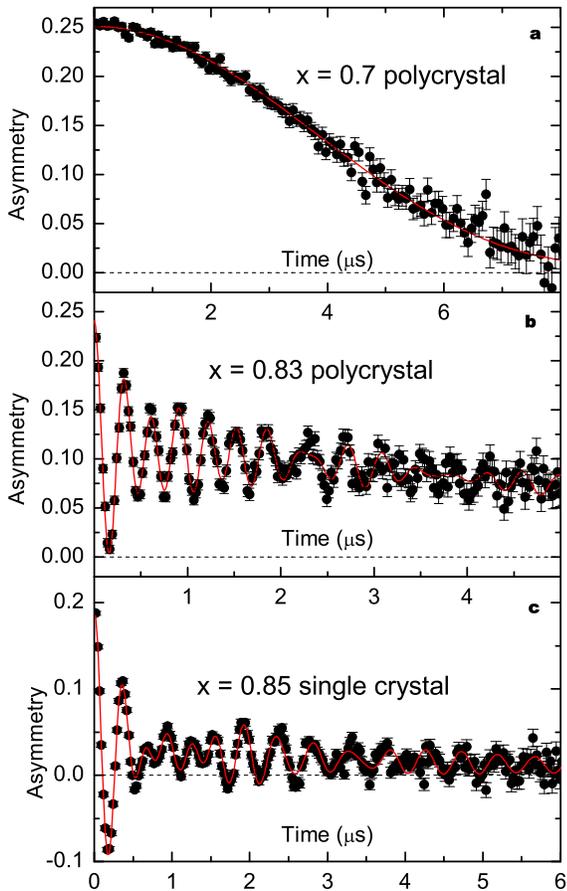}
\caption{\label{fig:usrmarkus1} Muon Spin Rotation ($\mathrm{\mu SR}$) data taken on \nax\ poly- and single crystals contrasting the low-T magnetic state for a sample with spin fluctuations and two with magnetic order. (a) The absence of any oscillatory contribution in the positron decay asymmetry corresponds to the absence of static magnetic LRO for a $x=0.7$ polycrystal. (b) In the $x=0.83$ polycrystal, the entire sample volume is magnetically ordered (within experimental limits of $\pm5\%$). (c) The same is true for the $x=0.85$ single crystal. Note that the saturation towards $1/3$ of the initial asymmetry value in panel (b) is due to the powder average but \textit{not} due to only partial magnetic order (statistically, $1/3$ of the internal fields have the same orientation as the muon spin polarisation in the polycrystals and thus do not force a muon precession).}
\end{figure}

\section{Results}

The results reveal several distinct regions in the phase diagram which will be discussed separately in the following.

\subsection{Spin fluctuation range ($\mathbf{x=0.6...0.75}$)}

The central observation in this region of the phase diagram is the absence of any magnetic long-range order (LRO), which can be deduced from the heat capacity data \cite{bruh06} in accordance with previous studies \cite{sak04}. Here we also show zero field muon-spin rotation spectra taken on $x=0.64$ and $x=0.7$ polycrystals. These $\mathrm{\mu SR}$ data to do not show any LRO down to $1.7\unit{K}$, which would result in an oscillatory part in the positron emission asymmetry (Fig.~\ref{fig:usrmarkus1}(a)). This is consistent with $\mathrm{\mu SR}$ results reported for similar compositions \cite{sug02,bruh06}. For a comprehensive description of the $\mathrm{\mu SR}$ technique applied to \nax\ see e.g.~Sugiyama et al.~\cite{sug03}. While $\mathrm{\mu SR}$ experiments reveal important information on the static magnetism they cannot elucidate on the possible presence of spin fluctuations with a time scale shorter than the order of $10^{-10} s$. On the other hand, from specific heat we get strong indications of low-energy magnetic excitations as will be discussed in the following.

\begin{figure}
\includegraphics[width=0.45\textwidth]{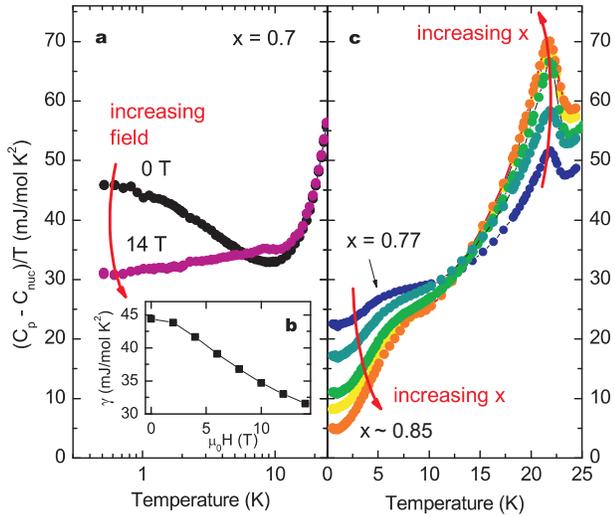}
\caption{\label{fig:markus1} Specific heat measured on \nax\ polycrystals highlighting the two distinct magnetic regions in the phase diagram. (a) The low-T DOS ($T<10\unit{K}$) is enhanced due to spin fluctuations for $0.6<x<0.75$. (b) Suppression of $C_{p}(T \rightarrow 0\unit{K})$ in an external magnetic field of up to $14\unit{T}$. (c) The onset of magnetic ordering for $x>0.75$ leads to a drastic reduction of the residual DOS ($T \rightarrow 0\unit{K}$) reflecting Fermi surface gapping. The nuclear contribution to the specific heat has been subtracted (see section III.B.2 and Fig.~4 for a discussion).}
\end{figure}

We note the absence of any sign of a phase transition in $C_{p}$ for $x=0.6...0.75$ which confirms the $\mathrm{\mu SR}$ results. However, the low temperature $C_{p}$ trend shows some peculiarity: $C_{p}$ distinctly increases below $T\approx 10K$, not inconsistent with a $T\cdot log(T)$ bahaviour, indicating an increasing excitation density at lowest temperatures (Fig.~\ref{fig:markus1}(a)). The contribution of the nuclei to the heat capacity has been subtracted for clarity (the nuclear heat capacity is discussed in section III.B.2). Dependent on $x$, values as large as $47\unit{mJ/mole \cdot K^{2}}$ are reached for $T \rightarrow 0\unit{K}$, consistent with other studies \cite{sak04}. In an external magnetic field, these additional excitations are significantly suppressed (Fig.~\ref{fig:markus1}(b)). Upon application of a $14\unit{T}$ external field, $\gamma(T \rightarrow 0K)$ decreases to $31\unit{mJ/mole \cdot K^{2}}$. In the $H$ range accessible with our magnet there is an indication of a saturation of $\gamma_{low}$ towards the value of $\gamma_{int}$, which is $25...30\unit{mJ/mole \cdot K^{2}}$ for $x=0.6...0.75$. Here (and also for higher x), $\gamma_{int}$ is significantly higher than in the $x<0.5$ range. This remarkable high DOS will be discussed in section IV.

Obviously, compared to $\gamma_{int}$ a strong renormalization of the electronic excitation density is present only at a low excitation energy scale ($\gamma_{low}$) which can be suppressed by magnetic fields. This behaviour is consistent with a magnetic origin of the apparent mass enhancement for $x = 0.6...0.75$ which is likely to be spin fluctuations that arise due to the proximity to magnetic ordering. This conclusion is in good agreement with NMR \cite{iha04} and neutron diffraction data \cite{hel06,bay05} that provide evidence for ferromagnetic in-plane fluctuations being present in \nax\ with $x = 0.7...0.75$. This conclusion has also been invoked before based on heat capacity studies \cite{sak04}, however without the evidence from field dependent measurements as those shown here. Additional support comes from theoretical studies yielding the onset of FM fluctuations for higher $x$ assisted by charge disproportionation \cite{lee04} or band structure effects \cite{korsh06}.

\subsection{Magnetic order range ($\mathbf{x\geq0.75}$)}

\subsubsection{Magnetic transitions}

For $x\geq0.75$, \nax\ develops a magnetically ordered groundstate as can be seen in the heat capacity and magnetization. For $x\geq0.8$ it affects the whole sample volume as shown by $\mathrm{\mu SR}$ data for polycrystals (Fig.~\ref{fig:usrmarkus1}(b)) and single crystals (Fig.~\ref{fig:usrmarkus1}(c)). For $x=0.75$, only $21\%$ magnetically ordered volume fraction has been reported \cite{sug03} which was interpreted as being the result of a phase separation into one magnetically ordered and one non-ordered phase \cite{sak04} to explain the large non-magnetic volume fraction. In single crystals the actual phase separation and thus coexistance of two distinct magnetic phases at low temperatures has been directly observed recently: In addition to the well-known $22\unit{K}$ AFM transition \cite{sal04,bay04,woo05,mot03} there is another transition at $8\unit{K}$ associated with a distinct second phase whose formation is dependent on the sample's cooling protocol \cite{schu07} and can be completely suppressed by thermal quenching. This points to sodium order as a driving force of magnetic order due to the patterned sodium Coulomb potential breaking the crystal symmetry and highlights the importance to account for multiple phases when interpreting low-T data. In polycrystalline samples (in which Na ordering effects are less pronounced in general) the $22\unit{K}$ SDW magnetic ordering can be observed and only a broad hump at lower temperatures suggests the $8\unit{K}$ magnetic ordering transition (Fig.~\ref{fig:markus1}(c) and Ref.\cite{sak04}). Even as the details of the development of magnetic LRO are somewhat different in single crystals, the $T \rightarrow 0\unit{K}$ behaviour is fully consistent with the polycrystals.

\begin{figure*}[ht]
\includegraphics[width=0.75\textwidth]{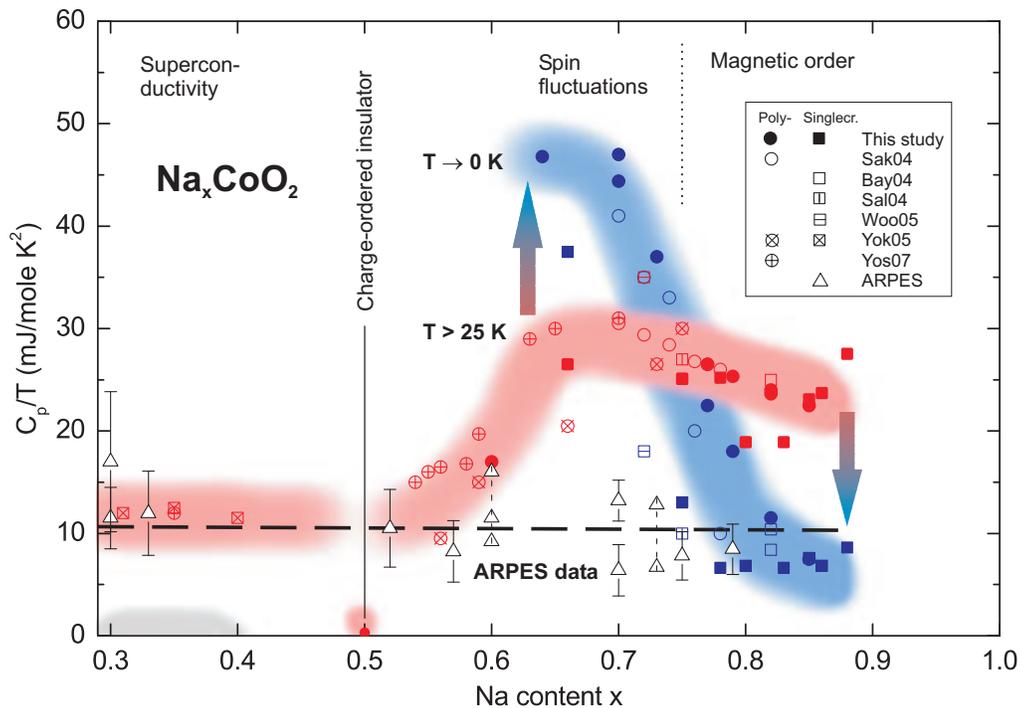}
\caption{\label{fig:phasediag} Phase diagram of the various electronic ground states in \naxs\ . The diagram contrasts the low energy electronic excitation density as measured via the specific heat at lowest temperatures (blue symbols) and intermediate temperatures (red symbols) with the ARPES-measured values (open triangles) and the approximate LDA band structure trend (dashed black line). Together with our data (filled symbols) numerous previous results by other groups are shown (open symbols). All specific heat data on single crystals are marked with square symbols while polycrystals are represented by circles. The colored arches are guides to the eye.}
\end{figure*}

Heat capacity measurements on $0.75<x<0.85$ polycrystalline samples show a systematic variation with $x$ (Fig.~\ref{fig:markus1}(c)): $\gamma_{low}$ decreases monotonically with $x$ while the intensity of the discontinuity at $22\unit{K}$ grows. Interestingly, within the experimental uncertainty of a few percent, $\gamma_{low}$ stays constant in fields up to $14\unit{T}$, in contrast to the $x=0.6...0.75$ range. The peak position of the discontinuities associated with the magnetic ordering transitions changes slightly with the applied magnetic field: The $8\unit{K}$ transition (only present in single crystals) moves up, the $22\unit{K}$ transition moves down, thus indicating a different nature of the magnetic state at $22\unit{K}$ and $8\unit{K}$ \cite{schu07}. Comparing $\gamma_{int}$ and $\gamma_{low}$ we note that $\gamma_{int}$ decreases slightly with $x$ while $\gamma_{low}$ falls dramatically from the highly renormalized value at $x=0.75$ and drops below the low-$x$ value of $\gamma_{int}$ of around $11\unit{mJ/mole\cdot K^{2}}$ at approximately $x=0.8$ (Fig.~\ref{fig:phasediag}). The reduction of $\gamma (T)$ from $\geq 20\unit{mJ/mole\cdot K^{2}}$ above the magnetic ordering to a value as low as $\approx 5\unit{mJ/mole\cdot K^{2}}$ for $\gamma_{low}$ reflects a substantial Fermi surface gapping ($75\%$), as expected for a SDW-like instability affecting a significant fraction of the Fermi surface(s). Thus our results are compatible with an at least partial Fermi surface gapping due to the onset of magnetic order in the high $x$ range.

\begin{figure}
\includegraphics[width=0.45\textwidth]{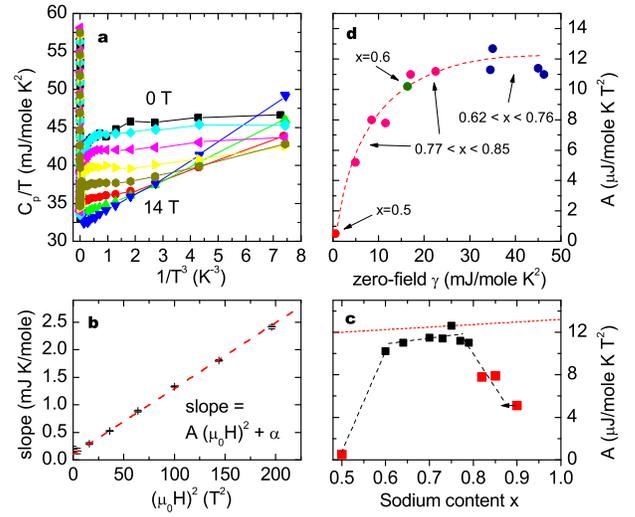}
\caption{\label{fig:nuccp} Nuclear specific heat analysis according to Eqns.~\ref{eq:cplowt} and \ref{eq:cplowta}. The main result is the reduction of the nuclear specific heat contribution (reflected in the coefficient A) in samples where (partial or complete) Fermi surface gapping is observed. The correspondance between the nuclear specific heat coefficient A and the DOS $\gamma$ is established in panel (d) with the comparison of the zero-field low-T $\gamma$ obtained from $C_{p}$ with A for several samples with different $x$. The dashed red line is a guide to the eye. Panel (c) shows the suppression of A for $x=0.5$ and $x>0.75$ compared to the expected value following Eqn.~\ref{eq:cplowta} (dashed red line). We consistently interpret this effect as Fermi surface gapping which reduces the thermal coupling of the Co and Na nuclei to the heat pulse used to probe the specific heat. The arrow accounts for the uncertainty of the sodium content for the sample with the highest $x$, which is likely to be sligthly smaller than the nominal $x$ (see section II). Panels (a) and (b) are auxiliary panels showing the experimental determination of A. Panel (a) is a $C_{p}/T$ vs. $1/T^{3}$ plot where $A\left(\mu_{0}H\right)^{2}+\alpha$ can be extracted as the slope. Plotting $A\left(\mu_{0}H\right)^{2}+\alpha$ vs. $\left(\mu_{0}H\right)^{2}$ yields A as the slope (Panel (b)). }
\end{figure}

\subsubsection{DOS and nuclear spin relaxation rate}

The reduction of DOS expressing itself in the small value of $\gamma_{low}$ as discussed above is consistently reflected in the low-temperature nuclear specific heat. The measured specific heat below $0.8\unit{K}$ is described by
\begin{equation}
C_{p}=\gamma_{low} T + A \left(\frac{\mu_{0}H}{T}\right)^{2}+\alpha \frac{1}{T^{2}},
\label{eq:cplowt}
\end{equation}
where the first term is the electronic heat capacity, the second term accounts for the nuclear heat capacity and the third and almost negligible term proportional to $T^{-2}$ is an unknown contribution which might be attributed to an electric quadrupole moment effect stemming from some localized Co $d$ electrons. The quality of the data lends itself to a quantitative analysis because all microscopic terms entering are known with the only free parameters being A and $\alpha$. The nuclear heat capacity originating from the $I=3/2$ Na and $I=7/2$ Co nuclei comprises the coefficient
\begin{equation}
A=N_{A}\frac{\mu_{N}^{2}}{k_{B}}\left[\frac{5}{4} x g_{Na}^{2}+\frac{21}{4}g_{Co}^{2}\right],
\label{eq:cplowta}
\end{equation}
with the g-factors $g_{Na}=1.478$ and $g_{Co}=1.322$, respectively. The lattice heat capacity is negligible at such low temperatures. Eqn.~\ref{eq:cplowt} yields a linear relationship of $C_{p}/T$ vs.~$T^{-3}$ (Fig.~\ref{fig:nuccp}(a)) from which $\gamma_{low}$ is determined as the ordinate intercept and $A\left(\mu_{0}H\right)^{2}+\alpha$ as the slope. Plotting the slope vs. $\mu_{0}H^{2}$ (Fig.~\ref{fig:nuccp}(b)) gives A and its variation with the sodium content $x$ is shown in Fig.~\ref{fig:nuccp}(c).

The variation of A as a function of $x$ is noteworthy for two reasons. For $x=0.6...0.8$, i.e.~in the range without long-range order (be it magnetic, or charge order at $x=0.5$ where $\gamma_{low}=0$), the value of A is very close to the expected value calculated from Eqn.~\ref{eq:cplowta} (dashed red line in Fig.~\ref{fig:nuccp}(c)). Remarkebly, A is much depressed or even unmeasurably small (for $x=0.5$) for these compositions where the Fermi surface is known from the previous analysis to be partly or fully gapped due to instabilities. A natural interpretation of these data takes into account that the nuclear spins obviously do not couple to the short heat pulse applied to the crystal in the $C_{p}$ measurement. Apparently, this coupling involves the DOS at $E_{F}$. Indeed, the values of A systematically vary with $\gamma_{low}(H=0)$ (Fig.~\ref{fig:nuccp}(d)).
This fact is also generally reflected in the much higher spin-lattice relaxation times of insulating materials compared to metals. Fig.~\ref{fig:nuccp} therefore provides independent additional self-consistent evidence for the FS gapping observed in the SDW state in addition to the well-known gapping for the charge-ordered insulator at $x=0.5$.

\section{Discussion}

\subsection{The low energy electronic excitation phase diagram}

The data just discussed are presented in an extended electronic excitation phase diagram (Fig.~\ref{fig:phasediag}). A highly consistent picture emerges when we combine our data with pertinent previously reported data. The open red and blue symbols represent the data from specific heat studies of Bayrakci et al. (Bay04,\cite{bay04}), Sales et al. (Sal04,\cite{sal04}), Sakurai et al. (Sak04,\cite{sak04}), Wooldridge et al. (Woo05,\cite{woo05}), Yokoi et al. (Yok05,\cite{yok05}) and Yoshizumi et al. (Yos07,\cite{yos07}). The open triangles stand for the ARPES data of Yang et al. \cite{yang04}, Hasan et al. \cite{has04}, Qian et al. \cite{qui06b,qui06c} and Geck et al. \cite{geck07} while the broken line marks the approximate values of the DOS as obtained by LDA calculations \cite{lee04,sin00}. In the high $x$ range, the two regions discussed above are clearly identified by distinctly different behaviour in terms of thermodynamic and magnetic properties. Inspecting the phase diagram, two aspects are worth noting:
\begin{enumerate}
	\item For $x>0.6$ the intermediate-temperature DOS $\gamma_{int}$ obtained at a low energy scale via the specific heat (red arch in Fig.~\ref{fig:phasediag}) increases by a factor 2..3 compared to the $x<0.5$ range and stays enhanced even for higher $x$ while both the LDA calculated $\gamma$ and the ARPES-measured high-energy DOS amount to around $10\unit{mJ/mole \cdot K^{2}}$ over the whole $x$ range (dashed line and triangles in Fig.~\ref{fig:phasediag}).
	\item The excitation density at lowest temperatures $\gamma_{low}$ (blue arch in Fig.~\ref{fig:phasediag}) is dominated by spin fluctuations for $x=0.6...0.75$ which lead to an additional enhancement compared to $\gamma_{int}$. However, due to magnetic order for $x>0.75$ the low-$T$ DOS is strongly reduced by the onset of Fermi surface gapping associated with the magnetic instabilities.
\end{enumerate}

\subsection{Structural issues}

The two distinct magnetic regions in the phase diagram, with spin fluctuations and with magnetic order, are separated by a transition region. The smooth transition between $x=0.7...0.8$ may well reflect the coexistence of more than one phase, as was first invoked by Sakurai et al. based on magnetization and heat capacity studies \cite{sak04b,sak04}. In the following it will be discussed in the light of extensive structural studies by Huang et al.\cite{hua04,hua04b,hua05}.
Structural transitions - particularly associated with the complex interplay between the temperature-dependent Na mobility and the inherent propensity in \nax\ to Na ordering - were intensely studied. There is evidence \cite{hua04} that only some well-defined compositions result in stable crystallographic phases that differ in the Na positioning in the unit cell and the occupation of the inequivalent Na sites. In between those particular $x$ values, phase separation into distinct phases with different Na content occurs \cite{hua04b,hua05}. This effect has been invoked in the high-$x$ range also to explain the presence of paramagnetic volume in otherwise magnetically ordered samples \cite{devau05,sak04}. Recently a particular Na rearrangement process was shown to lead to distinct magnetically ordered states associated with different phases in the same crystal and the connection to Na diffusion as its origin was established \cite{schu07}.
In this context the distinct but gradual change observed in $\gamma_{int}$ and $\gamma_{low}$ for $x=0.7...0.8$ which marks the transition between the two adjacent magnetic ranges could well be the result of the coexistence of one high-$x$ phase hosting magnetic order and another phase with lower $x$ associated with the spin fluctuations. Interestingly, in the studies by Huang et al.~\cite{hua04} a coexistance between the so called H1 and H2 phases is observed in this narrow $x$ range. Thus it would be reasonable to associate the spin fluctuation phase with H1 and the SDW phase with H2. It would be worthwile, however, to clarify in detail to what degree structural parameters, other than the ones distinguishing H1, H2 and H3, need to be included in order to capture the subtle differences of the various phases. 

\subsection{Comparing $\mathbf{C_{p}}$, ARPES and band structure calculations}

As shown in the phase diagram above, in addition to the spin fluctuation contribution to $\gamma$ that vanishes in high magnetic fields, there is another enhancement for $x>0.6$ that is not suppressed by magnetic fields and is also detectible at intermediate temperatures ($T>25\unit{K}$). In the following paragraph we concentrate our discussion on $\gamma_{int}$ as it is not affected by spin fluctuations or magnetic order. We note that $\gamma_{int}$ is enhanced by a factor of 2..3 compared to the expected values from band structure calculations (around $10\unit{mJ/mole\cdot K^{2}}$)\cite{lee04,sin00} and the ARPES experimental data ($8...15\unit{mJ/mole\cdot K^{2}}$). In addition to our findings, two other extensive studies \cite{yos07,yok05} support our results.

Obviously the electronic DOS is significantly higher for $x>0.6$ which is tentatively ascribed to an additional band crossing the Fermi surface following Yoshizumi et al. \cite{yos07}. This could most likely be the $a_{1g}$ character band which has been calculated to develop a dip near the $\Gamma$ point for higher $x$ \cite{korsh06}. This would result in the emergence of an electron pocket when $E_{F}$ touches the band from below upon filling of the Co-O states with increasing x. Consistent with our findings, the onset of small-$\mathbf{Q}$ magnetic fluctuations is predicted to accompany the changes of the Fermi surface topology \cite{korsh06}.
A second explanation would involve the $e'_{g}$ character hole pockets that were intensely sought for in numerous ARPES studies.
In a rigid band picture these should rather disappear with increasing Fermi level which has also been predicted by numerous theoretical studies \cite{geck07,zhou05} and recently been claimed to be observed experimentally \cite{lav07}. Even if the hole pockets were stabilized also for higher $x$ as predicted by some models \cite{joh04,ish05}, the DOS behaviour observed here could hardly be explained.
Further, another possibility would be ascribing the observed enhancement for higher $x$ purely to the onset of strong electron-electron correlations in which case a significant impact on the band structure was anticipated by theoretical studies \cite{lee04,joh04} leading to a band picture being more consistent with the ARPES results. However, no origin for such a dramatic increase of correlations around $x\approx 0.6$ could be identified so far rendering this option rather speculative.

It is interesting to note that $\gamma$ calculated from ARPES-measured band dispersions involving one major Fermi surface agrees well with the LDA-calculated $\gamma$ across the whole phase diagram although there are severe discrepancies between the ARPES findings and the calculated dispersions:
The bandwidth observed in the ARPES studies is systematically reduced compared to bandstructure calculations, being $>100\unit{meV}$ instead of $1...1.4\unit{eV}$. This leads to the conclusion of the hopping integral $t$ being renormalized by a factor $10...15$ already at higher temperatures which should be reflected in a dramatically enhanced DOS compared to the calculations. However, the absence of the $e'_{g}$ character hole pockets or other additional Fermi surface segments seem to exactly compensate this effect in the ARPES measurements.
Thus it appears that the ARPES results so far do not account for the highly enhanced DOS values for $x>0.6$. Even though it is commonly noted that there must be strong correlations present to explain the small bandwidth, in combination with the experimentally observed Fermi surface topology this does not suffice to explain the high $\gamma$ values.

Interestingly, there is a discrepancy concerning the dimensionality of the electronic system in the high $x$ range among the ARPES studies. One study sees a significant $z$-dispersion resulting in additional FS segments \cite{qui06c} while in most cases the system seems to remain quasi-2D \cite{geck07,has04}. If additional Fermi surface caused by three-dimensionality was present, a higher DOS is consequently to be expected.

To decide which of these scenarios - or possibly others - realistically describe the electronic states in \nax\ is an intriguing puzzle and awaits clarification by further studies. In the light of the recently observed coupling between the Fermi surface instabilities and the degree of freedom imposed by sodium mobility and the plethora of possible ordering patterns \cite{rog07,zand04,schu07}, it is obvious that future theoretical studies should reach beyond the virtual crystal approximation and realistically model the interplay between Na driven Coulomb potential patterns, Fermi surface topology \& dimensionality and charge/magnetic order. 

\section{Conclusion}

The interpretation of our heat capacity and muon-spin rotation data provides a consistent picture of the \nax\ low energy electronic excitation density for both intermediate and low temperatures over a large sodium concentration range and for both single and polycrystals. The comparison with previous data as compiled in Fig.~\ref{fig:phasediag} provides a comprehensive summary of the current and previously published data and reflects the agreement among the results from various studies. Some of the conclusions arrived at in the present study have been suggested before, such as the coexistence of a spin fluctuation and a magnetically ordered phase for $0.7<x<0.78$ \cite{sak04}. New insight, however, is gained from our in-depth analysis of both $\gamma_{int}$ and $\gamma_{low}$: Our field-dependent heat capacity measurements in quantitative comparison with ARPES and LDA results reveal a rich and complex picture which renders spin fluctuations to be only one out of two coexisting low energy excitation density enhancement mechanisms for $x>0.6$. 

For the whole $x$ range the \textit{high energy} band dispersion as measured by ARPES yields a density of states consistent with LDA band structure calculations (DOS around $10 \unit{mJ/mole\cdot K^{2}}$). However, the \textit{low energy} excitation spectrum ($<100\unit{K}$) as measured by specific heat reveals distinctly richer phenomena: For $x>0.6$ the intermediate-$T$ DOS is 2...3 times higher ($25...30 \unit{mJ/mole\cdot K^{2}}$). In the range $0.6<x<0.75$ the low-temperature excitation density is dominated by spin fluctuations that can be suppressed in a magnetic field and cause even stronger renormalization of $\gamma$ as compared to the intermediate-temperature range, but no magnetic long-range order is present. A variety of magnetically ordered states is then observed for $x>0.75$, including a $8\unit{K}$ ferromagnetic-type transition in addition to the previously studied SDW-like $22\unit{K}$ transition.

In the light of the low DOS derived from ARPES and LDA calculations and the existence of a second DOS enhancement mechanism seen in the excitation density phase diagram (be it a band structure or renormalization effect), the question whether Fermi surface gapping caused by magnetic ordering is present in the high-$x$ range is not trivially answered by a loss of low-$T$ DOS as compared to intermediate-$T$ DOS. Here, our nuclear heat capacity analysis provides important additional evidence based upon which it can be concluded that the magnetic order involving the itinerant electrons indeed causes the disappearance of part of the Fermi surface.

Thus, the $x > 0.6$ range of the phase diagram is highlighted to be most interesting by the presence of a high DOS and Fermi surface loss through magnetic instabilities. The revisited phase diagram illustrates the rich physics of the correlated electrons moving on the triangular lattice in the presence of a controllable potential landscape defined by ordering patterns formed by highly mobile Na ions.

This work is partly based on muon experiments performed at the Swiss Muon Source, Paul Scherrer Institut (PSI), Villigen (Switzerland).

\begin{acknowledgments}
We gratefully acknowledge helpful discussions with M. Sigrist, and partial financial support by the Swiss National Science Foundation through the NCCR MaNEP.
\end{acknowledgments}


\bibliographystyle{apsrev}

\begin{thebibliography}{40}
\expandafter\ifx\csname natexlab\endcsname\relax\def\natexlab#1{#1}\fi
\expandafter\ifx\csname bibnamefont\endcsname\relax
  \def\bibnamefont#1{#1}\fi
\expandafter\ifx\csname bibfnamefont\endcsname\relax
  \def\bibfnamefont#1{#1}\fi
\expandafter\ifx\csname citenamefont\endcsname\relax
  \def\citenamefont#1{#1}\fi
\expandafter\ifx\csname url\endcsname\relax
  \def\url#1{\texttt{#1}}\fi
\expandafter\ifx\csname urlprefix\endcsname\relax\def\urlprefix{URL }\fi
\providecommand{\bibinfo}[2]{#2}
\providecommand{\eprint}[2][]{\url{#2}}

\bibitem[{\citenamefont{Terasaki et~al.}(1997)\citenamefont{Terasaki, Sasago,
  and Uchinokura}}]{ter97}
\bibinfo{author}{\bibfnamefont{I.}~\bibnamefont{Terasaki}},
  \bibinfo{author}{\bibfnamefont{Y.}~\bibnamefont{Sasago}}, \bibnamefont{and}
  \bibinfo{author}{\bibfnamefont{K.}~\bibnamefont{Uchinokura}},
  \bibinfo{journal}{Phys.\ Rev.\ B} \textbf{\bibinfo{volume}{56}},
  \bibinfo{pages}{R12685} (\bibinfo{year}{1997}).

\bibitem[{\citenamefont{Lee et~al.}(2006)\citenamefont{Lee, Viciu, Li, Wang,
  Foo, Watauchi, Pascal, Cava, and Ong}}]{lee06}
\bibinfo{author}{\bibfnamefont{M.}~\bibnamefont{Lee}},
  \bibinfo{author}{\bibfnamefont{L.}~\bibnamefont{Viciu}},
  \bibinfo{author}{\bibfnamefont{L.}~\bibnamefont{Li}},
  \bibinfo{author}{\bibfnamefont{Y.~Y.} \bibnamefont{Wang}},
  \bibinfo{author}{\bibfnamefont{M.~L.} \bibnamefont{Foo}},
  \bibinfo{author}{\bibfnamefont{S.}~\bibnamefont{Watauchi}},
  \bibinfo{author}{\bibfnamefont{R.~A.} \bibnamefont{Pascal}},
  \bibinfo{author}{\bibfnamefont{R.~J.} \bibnamefont{Cava}}, \bibnamefont{and}
  \bibinfo{author}{\bibfnamefont{N.~P.} \bibnamefont{Ong}},
  \bibinfo{journal}{Nat.\ Mater.} \textbf{\bibinfo{volume}{5}},
  \bibinfo{pages}{537} (\bibinfo{year}{2006}).

\bibitem[{\citenamefont{Takada et~al.}(2003)\citenamefont{Takada, Sakurai,
  Takayama-Muromachi, Izumi, Dilanian, and Sasaki}}]{tak03}
\bibinfo{author}{\bibfnamefont{K.}~\bibnamefont{Takada}},
  \bibinfo{author}{\bibfnamefont{H.}~\bibnamefont{Sakurai}},
  \bibinfo{author}{\bibfnamefont{E.}~\bibnamefont{Takayama-Muromachi}},
  \bibinfo{author}{\bibfnamefont{F.}~\bibnamefont{Izumi}},
  \bibinfo{author}{\bibfnamefont{R.~A.} \bibnamefont{Dilanian}},
  \bibnamefont{and} \bibinfo{author}{\bibfnamefont{T.}~\bibnamefont{Sasaki}},
  \bibinfo{journal}{Nature} \textbf{\bibinfo{volume}{422}}, \bibinfo{pages}{53}
  (\bibinfo{year}{2003}).

\bibitem[{\citenamefont{Foo et~al.}(2004)\citenamefont{Foo, Wang, Watauchi,
  Zandbergen, He, Cava, and Ong}}]{foo04}
\bibinfo{author}{\bibfnamefont{M.~L.} \bibnamefont{Foo}},
  \bibinfo{author}{\bibfnamefont{Y.~Y.} \bibnamefont{Wang}},
  \bibinfo{author}{\bibfnamefont{S.}~\bibnamefont{Watauchi}},
  \bibinfo{author}{\bibfnamefont{H.~W.} \bibnamefont{Zandbergen}},
  \bibinfo{author}{\bibfnamefont{T.}~\bibnamefont{He}},
  \bibinfo{author}{\bibfnamefont{R.~J.} \bibnamefont{Cava}}, \bibnamefont{and}
  \bibinfo{author}{\bibfnamefont{N.~P.} \bibnamefont{Ong}},
  \bibinfo{journal}{Phys.\ Rev.\ Lett.} \textbf{\bibinfo{volume}{92}},
  \bibinfo{pages}{247001} (\bibinfo{year}{2004}).

\bibitem[{\citenamefont{Boothroyd et~al.}(2004)\citenamefont{Boothroyd, Coldea,
  Tennant, Prabhakaran, Helme, and Frost}}]{boo04}
\bibinfo{author}{\bibfnamefont{A.~T.} \bibnamefont{Boothroyd}},
  \bibinfo{author}{\bibfnamefont{R.}~\bibnamefont{Coldea}},
  \bibinfo{author}{\bibfnamefont{D.~A.} \bibnamefont{Tennant}},
  \bibinfo{author}{\bibfnamefont{D.}~\bibnamefont{Prabhakaran}},
  \bibinfo{author}{\bibfnamefont{L.~M.} \bibnamefont{Helme}}, \bibnamefont{and}
  \bibinfo{author}{\bibfnamefont{C.~D.} \bibnamefont{Frost}},
  \bibinfo{journal}{Phys.\ Rev.\ Lett.} \textbf{\bibinfo{volume}{92}},
  \bibinfo{pages}{197201} (\bibinfo{year}{2004}).

\bibitem[{\citenamefont{Sales et~al.}(2004)\citenamefont{Sales, Jin, Affholter,
  Khalifah, Veith, and Mandrus}}]{sal04}
\bibinfo{author}{\bibfnamefont{B.~C.} \bibnamefont{Sales}},
  \bibinfo{author}{\bibfnamefont{R.}~\bibnamefont{Jin}},
  \bibinfo{author}{\bibfnamefont{K.~A.} \bibnamefont{Affholter}},
  \bibinfo{author}{\bibfnamefont{P.}~\bibnamefont{Khalifah}},
  \bibinfo{author}{\bibfnamefont{G.~M.} \bibnamefont{Veith}}, \bibnamefont{and}
  \bibinfo{author}{\bibfnamefont{D.}~\bibnamefont{Mandrus}},
  \bibinfo{journal}{Phys.\ Rev.\ B} \textbf{\bibinfo{volume}{70}},
  \bibinfo{pages}{174419} (\bibinfo{year}{2004}).

\bibitem[{\citenamefont{Bayrakci et~al.}(2004)\citenamefont{Bayrakci, Bernhard,
  Chen, Keimer, Kremer, Lemmens, Lin, Niedermayer, and Strempfer}}]{bay04}
\bibinfo{author}{\bibfnamefont{S.~P.} \bibnamefont{Bayrakci}},
  \bibinfo{author}{\bibfnamefont{C.}~\bibnamefont{Bernhard}},
  \bibinfo{author}{\bibfnamefont{D.~P.} \bibnamefont{Chen}},
  \bibinfo{author}{\bibfnamefont{B.}~\bibnamefont{Keimer}},
  \bibinfo{author}{\bibfnamefont{R.~K.} \bibnamefont{Kremer}},
  \bibinfo{author}{\bibfnamefont{P.}~\bibnamefont{Lemmens}},
  \bibinfo{author}{\bibfnamefont{C.~T.} \bibnamefont{Lin}},
  \bibinfo{author}{\bibfnamefont{C.}~\bibnamefont{Niedermayer}},
  \bibnamefont{and}
  \bibinfo{author}{\bibfnamefont{J.}~\bibnamefont{Strempfer}},
  \bibinfo{journal}{Phys.\ Rev.\ B} \textbf{\bibinfo{volume}{69}},
  \bibinfo{pages}{100410(R)} (\bibinfo{year}{2004}).

\bibitem[{\citenamefont{Wooldridge et~al.}(2005)\citenamefont{Wooldridge, Paul,
  Balakrishnan, and Lees}}]{woo05}
\bibinfo{author}{\bibfnamefont{J.}~\bibnamefont{Wooldridge}},
  \bibinfo{author}{\bibfnamefont{D.~M.} \bibnamefont{Paul}},
  \bibinfo{author}{\bibfnamefont{G.}~\bibnamefont{Balakrishnan}},
  \bibnamefont{and} \bibinfo{author}{\bibfnamefont{M.~R.} \bibnamefont{Lees}},
  \bibinfo{journal}{J.\ Phys-Condens.\ Mat.} \textbf{\bibinfo{volume}{17}},
  \bibinfo{pages}{707} (\bibinfo{year}{2005}).

\bibitem[{\citenamefont{Motohashi et~al.}(2001)\citenamefont{Motohashi,
  Naujalis, Ueda, Isawa, Karppinen, and Yamauchi}}]{mot01}
\bibinfo{author}{\bibfnamefont{T.}~\bibnamefont{Motohashi}},
  \bibinfo{author}{\bibfnamefont{E.}~\bibnamefont{Naujalis}},
  \bibinfo{author}{\bibfnamefont{R.}~\bibnamefont{Ueda}},
  \bibinfo{author}{\bibfnamefont{K.}~\bibnamefont{Isawa}},
  \bibinfo{author}{\bibfnamefont{M.}~\bibnamefont{Karppinen}},
  \bibnamefont{and} \bibinfo{author}{\bibfnamefont{H.}~\bibnamefont{Yamauchi}},
  \bibinfo{journal}{Applied Physics Letters} \textbf{\bibinfo{volume}{79}},
  \bibinfo{pages}{1480} (\bibinfo{year}{2001}).

\bibitem[{\citenamefont{Prabhakaran et~al.}(2004)\citenamefont{Prabhakaran,
  Boothroyd, Coldea, and Charnley}}]{prab04}
\bibinfo{author}{\bibfnamefont{D.}~\bibnamefont{Prabhakaran}},
  \bibinfo{author}{\bibfnamefont{A.~T.} \bibnamefont{Boothroyd}},
  \bibinfo{author}{\bibfnamefont{R.}~\bibnamefont{Coldea}}, \bibnamefont{and}
  \bibinfo{author}{\bibfnamefont{N.~R.} \bibnamefont{Charnley}},
  \bibinfo{journal}{J.\ Cryst.\ Growth} \textbf{\bibinfo{volume}{271}},
  \bibinfo{pages}{74} (\bibinfo{year}{2004}).

\bibitem[{\citenamefont{Bayrakci et~al.}(2005)\citenamefont{Bayrakci, Mirebeau,
  Bourges, Sidis, Enderle, Mesot, Chen, Lin, and Keimer}}]{bay05}
\bibinfo{author}{\bibfnamefont{S.~P.} \bibnamefont{Bayrakci}},
  \bibinfo{author}{\bibfnamefont{I.}~\bibnamefont{Mirebeau}},
  \bibinfo{author}{\bibfnamefont{P.}~\bibnamefont{Bourges}},
  \bibinfo{author}{\bibfnamefont{Y.}~\bibnamefont{Sidis}},
  \bibinfo{author}{\bibfnamefont{M.}~\bibnamefont{Enderle}},
  \bibinfo{author}{\bibfnamefont{J.}~\bibnamefont{Mesot}},
  \bibinfo{author}{\bibfnamefont{D.~P.} \bibnamefont{Chen}},
  \bibinfo{author}{\bibfnamefont{C.~T.} \bibnamefont{Lin}}, \bibnamefont{and}
  \bibinfo{author}{\bibfnamefont{B.}~\bibnamefont{Keimer}},
  \bibinfo{journal}{Phys.\ Rev.\ Lett.} \textbf{\bibinfo{volume}{94}},
  \bibinfo{pages}{157205} (\bibinfo{year}{2005}).

\bibitem[{\citenamefont{Helme et~al.}(2006)\citenamefont{Helme, Boothroyd,
  Coldea, Prabhakaran, Stunault, McIntyre, and Kernavanois}}]{hel06}
\bibinfo{author}{\bibfnamefont{L.~M.} \bibnamefont{Helme}},
  \bibinfo{author}{\bibfnamefont{A.~T.} \bibnamefont{Boothroyd}},
  \bibinfo{author}{\bibfnamefont{R.}~\bibnamefont{Coldea}},
  \bibinfo{author}{\bibfnamefont{D.}~\bibnamefont{Prabhakaran}},
  \bibinfo{author}{\bibfnamefont{A.}~\bibnamefont{Stunault}},
  \bibinfo{author}{\bibfnamefont{G.~J.} \bibnamefont{McIntyre}},
  \bibnamefont{and}
  \bibinfo{author}{\bibfnamefont{N.}~\bibnamefont{Kernavanois}},
  \bibinfo{journal}{Phys.\ Rev.\ B} \textbf{\bibinfo{volume}{73}},
  \bibinfo{pages}{054405} (\bibinfo{year}{2006}).

\bibitem[{\citenamefont{Johannes et~al.}(2004)\citenamefont{Johannes,
  Papaconstantopoulos, Singh, and Mehl}}]{joh04}
\bibinfo{author}{\bibfnamefont{M.~D.} \bibnamefont{Johannes}},
  \bibinfo{author}{\bibfnamefont{D.~A.} \bibnamefont{Papaconstantopoulos}},
  \bibinfo{author}{\bibfnamefont{D.~J.} \bibnamefont{Singh}}, \bibnamefont{and}
  \bibinfo{author}{\bibfnamefont{M.~J.} \bibnamefont{Mehl}},
  \bibinfo{journal}{Europhys.\ Lett.} \textbf{\bibinfo{volume}{68}},
  \bibinfo{pages}{433} (\bibinfo{year}{2004}).

\bibitem[{\citenamefont{Geck et~al.}(2007)\citenamefont{Geck, Borisenko,
  Berger, Eschrig, Fink, Knupfer, Koepernik, Koitzsch, Kordyuk, Zabolotnyy
  et~al.}}]{geck07}
\bibinfo{author}{\bibfnamefont{J.}~\bibnamefont{Geck}},
  \bibinfo{author}{\bibfnamefont{S.~V.} \bibnamefont{Borisenko}},
  \bibinfo{author}{\bibfnamefont{H.}~\bibnamefont{Berger}},
  \bibinfo{author}{\bibfnamefont{H.}~\bibnamefont{Eschrig}},
  \bibinfo{author}{\bibfnamefont{J.}~\bibnamefont{Fink}},
  \bibinfo{author}{\bibfnamefont{M.}~\bibnamefont{Knupfer}},
  \bibinfo{author}{\bibfnamefont{K.}~\bibnamefont{Koepernik}},
  \bibinfo{author}{\bibfnamefont{A.}~\bibnamefont{Koitzsch}},
  \bibinfo{author}{\bibfnamefont{A.~A.} \bibnamefont{Kordyuk}},
  \bibinfo{author}{\bibfnamefont{V.~B.} \bibnamefont{Zabolotnyy}},
  \bibnamefont{et~al.}, \bibinfo{journal}{Phys.\ Rev.\ Lett.}
  \textbf{\bibinfo{volume}{99}}, \bibinfo{pages}{046403}
  (\bibinfo{year}{2007}).

\bibitem[{\citenamefont{Qian et~al.}(2006{\natexlab{a}})\citenamefont{Qian,
  Wray, Hsieh, Viciu, Cava, Luo, Wu, Wang, and Hasan}}]{qui06c}
\bibinfo{author}{\bibfnamefont{D.}~\bibnamefont{Qian}},
  \bibinfo{author}{\bibfnamefont{L.}~\bibnamefont{Wray}},
  \bibinfo{author}{\bibfnamefont{D.}~\bibnamefont{Hsieh}},
  \bibinfo{author}{\bibfnamefont{L.}~\bibnamefont{Viciu}},
  \bibinfo{author}{\bibfnamefont{R.~J.} \bibnamefont{Cava}},
  \bibinfo{author}{\bibfnamefont{J.~L.} \bibnamefont{Luo}},
  \bibinfo{author}{\bibfnamefont{D.}~\bibnamefont{Wu}},
  \bibinfo{author}{\bibfnamefont{N.~L.} \bibnamefont{Wang}}, \bibnamefont{and}
  \bibinfo{author}{\bibfnamefont{M.~Z.} \bibnamefont{Hasan}},
  \bibinfo{journal}{Phys.\ Rev.\ Lett.} \textbf{\bibinfo{volume}{97}},
  \bibinfo{pages}{186405} (\bibinfo{year}{2006}{\natexlab{a}}).

\bibitem[{\citenamefont{Yang et~al.}(2004)\citenamefont{Yang, Wang, Sekharan,
  Matsui, Souma, Sato, Takahashi, Takeuchi, Campuzano, Jin et~al.}}]{yang04}
\bibinfo{author}{\bibfnamefont{H.~B.} \bibnamefont{Yang}},
  \bibinfo{author}{\bibfnamefont{S.~C.} \bibnamefont{Wang}},
  \bibinfo{author}{\bibfnamefont{A.~K.~P.} \bibnamefont{Sekharan}},
  \bibinfo{author}{\bibfnamefont{H.}~\bibnamefont{Matsui}},
  \bibinfo{author}{\bibfnamefont{S.}~\bibnamefont{Souma}},
  \bibinfo{author}{\bibfnamefont{T.}~\bibnamefont{Sato}},
  \bibinfo{author}{\bibfnamefont{T.}~\bibnamefont{Takahashi}},
  \bibinfo{author}{\bibfnamefont{T.}~\bibnamefont{Takeuchi}},
  \bibinfo{author}{\bibfnamefont{J.~C.} \bibnamefont{Campuzano}},
  \bibinfo{author}{\bibfnamefont{R.}~\bibnamefont{Jin}}, \bibnamefont{et~al.},
  \bibinfo{journal}{Phys.\ Rev.\ Lett.} \textbf{\bibinfo{volume}{92}},
  \bibinfo{pages}{246403} (\bibinfo{year}{2004}).

\bibitem[{\citenamefont{Lynn et~al.}(2003)\citenamefont{Lynn, Huang, Brown,
  Miller, Foo, Schaak, Jones, Mackey, and Cava}}]{lynn03}
\bibinfo{author}{\bibfnamefont{J.~W.} \bibnamefont{Lynn}},
  \bibinfo{author}{\bibfnamefont{Q.}~\bibnamefont{Huang}},
  \bibinfo{author}{\bibfnamefont{C.~M.} \bibnamefont{Brown}},
  \bibinfo{author}{\bibfnamefont{V.~L.} \bibnamefont{Miller}},
  \bibinfo{author}{\bibfnamefont{M.~L.} \bibnamefont{Foo}},
  \bibinfo{author}{\bibfnamefont{R.~E.} \bibnamefont{Schaak}},
  \bibinfo{author}{\bibfnamefont{C.~Y.} \bibnamefont{Jones}},
  \bibinfo{author}{\bibfnamefont{E.~A.} \bibnamefont{Mackey}},
  \bibnamefont{and} \bibinfo{author}{\bibfnamefont{R.~J.} \bibnamefont{Cava}},
  \bibinfo{journal}{Physical Review B} \textbf{\bibinfo{volume}{68}},
  \bibinfo{pages}{214516} (\bibinfo{year}{2003}).

\bibitem[{\citenamefont{Sugiyama et~al.}(2002)\citenamefont{Sugiyama, Itahara,
  Tani, Brewer, and Ansaldo}}]{sug02}
\bibinfo{author}{\bibfnamefont{J.}~\bibnamefont{Sugiyama}},
  \bibinfo{author}{\bibfnamefont{H.}~\bibnamefont{Itahara}},
  \bibinfo{author}{\bibfnamefont{T.}~\bibnamefont{Tani}},
  \bibinfo{author}{\bibfnamefont{J.~H.} \bibnamefont{Brewer}},
  \bibnamefont{and} \bibinfo{author}{\bibfnamefont{E.~J.}
  \bibnamefont{Ansaldo}}, \bibinfo{journal}{Physical Review B}
  \textbf{\bibinfo{volume}{66}}, \bibinfo{pages}{134413}
  (\bibinfo{year}{2002}).

\bibitem[{\citenamefont{Br\"uhwiler et~al.}(2006)\citenamefont{Br\"uhwiler,
  Batlogg, Kazakov, Niedermayer, and Karpinski}}]{bruh06}
\bibinfo{author}{\bibfnamefont{M.}~\bibnamefont{Br\"uwiler}},
  \bibinfo{author}{\bibfnamefont{B.}~\bibnamefont{Batlogg}},
  \bibinfo{author}{\bibfnamefont{S.~M.} \bibnamefont{Kazakov}},
  \bibinfo{author}{\bibfnamefont{C.}~\bibnamefont{Niedermayer}},
  \bibnamefont{and}
  \bibinfo{author}{\bibfnamefont{J.}~\bibnamefont{Karpinski}},
  \bibinfo{journal}{Physica B} \textbf{\bibinfo{volume}{378-380}},
  \bibinfo{pages}{630} (\bibinfo{year}{2006}).

\bibitem[{\citenamefont{Sugiyama et~al.}(2003)\citenamefont{Sugiyama, Itahara,
  Brewer, Ansaldo, Motohashi, Karppinen, and Yamauchi}}]{sug03}
\bibinfo{author}{\bibfnamefont{J.}~\bibnamefont{Sugiyama}},
  \bibinfo{author}{\bibfnamefont{H.}~\bibnamefont{Itahara}},
  \bibinfo{author}{\bibfnamefont{J.~H.} \bibnamefont{Brewer}},
  \bibinfo{author}{\bibfnamefont{E.~J.} \bibnamefont{Ansaldo}},
  \bibinfo{author}{\bibfnamefont{T.}~\bibnamefont{Motohashi}},
  \bibinfo{author}{\bibfnamefont{M.}~\bibnamefont{Karppinen}},
  \bibnamefont{and} \bibinfo{author}{\bibfnamefont{H.}~\bibnamefont{Yamauchi}},
  \bibinfo{journal}{Phys.\ Rev.\ B} \textbf{\bibinfo{volume}{67}},
  \bibinfo{pages}{214420} (\bibinfo{year}{2003}).

\bibitem[{\citenamefont{Sakurai et~al.}(2004)\citenamefont{Sakurai, Tsujii, and Takayama-Muromachi}}]{sak04}
\bibinfo{author}{\bibfnamefont{H.}~\bibnamefont{Sakurai}},
  \bibinfo{author}{\bibfnamefont{N.}~\bibnamefont{Tsujii}}, \bibnamefont{and}
  \bibinfo{author}{\bibfnamefont{E.}~\bibnamefont{Takayama-Muromachi}},
  \bibinfo{journal}{J.\ Phys.\ Soc.\ Jpn} \textbf{\bibinfo{volume}{73}},
  \bibinfo{pages}{2393} (\bibinfo{year}{2004}).
  
\bibitem[{\citenamefont{Sakurai et~al.}(2004)\citenamefont{Sakurai, Takenouchi,
  Tsujii, and Takayama-Muromachi}}]{sak04b}
\bibinfo{author}{\bibfnamefont{H.}~\bibnamefont{Sakurai}},
  \bibinfo{author}{\bibfnamefont{S.}~\bibnamefont{Takenouchi}},
  \bibinfo{author}{\bibfnamefont{N.}~\bibnamefont{Tsujii}}, \bibnamefont{and}
  \bibinfo{author}{\bibfnamefont{E.}~\bibnamefont{Takayama-Muromachi}},
  \bibinfo{journal}{J.\ Phys.\ Soc.\ Jpn} \textbf{\bibinfo{volume}{73}},
  \bibinfo{pages}{2081} (\bibinfo{year}{2004}).

\bibitem[{\citenamefont{Ihara et~al.}(2004)\citenamefont{Ihara, Ishida,
  Michioka, Kato, Yoshimura, Sakurai, and Takayama-Muromachi}}]{iha04}
\bibinfo{author}{\bibfnamefont{Y.}~\bibnamefont{Ihara}},
  \bibinfo{author}{\bibfnamefont{K.}~\bibnamefont{Ishida}},
  \bibinfo{author}{\bibfnamefont{C.}~\bibnamefont{Michioka}},
  \bibinfo{author}{\bibfnamefont{M.}~\bibnamefont{Kato}},
  \bibinfo{author}{\bibfnamefont{K.}~\bibnamefont{Yoshimura}},
  \bibinfo{author}{\bibfnamefont{H.}~\bibnamefont{Sakurai}}, \bibnamefont{and}
  \bibinfo{author}{\bibfnamefont{E.}~\bibnamefont{Takayama-Muromachi}},
  \bibinfo{journal}{J.~Phys.~Soc.~Jpn.} \textbf{\bibinfo{volume}{73}},
  \bibinfo{pages}{2963} (\bibinfo{year}{2004}).

\bibitem[{\citenamefont{Lee et~al.}(2004)\citenamefont{Lee, Kunes, and
  Pickett}}]{lee04}
\bibinfo{author}{\bibfnamefont{K.~W.} \bibnamefont{Lee}},
  \bibinfo{author}{\bibfnamefont{J.}~\bibnamefont{Kunes}}, \bibnamefont{and}
  \bibinfo{author}{\bibfnamefont{W.~E.} \bibnamefont{Pickett}},
  \bibinfo{journal}{Phys.\ Rev.\ B} \textbf{\bibinfo{volume}{70}},
  \bibinfo{pages}{045104} (\bibinfo{year}{2004}).

\bibitem[{\citenamefont{Korshunov et~al.}(2006)\citenamefont{Korshunov, Eremin,
  Shorikov, and Anisimov}}]{korsh06}
\bibinfo{author}{\bibfnamefont{M.~M.} \bibnamefont{Korshunov}},
  \bibinfo{author}{\bibfnamefont{I.}~\bibnamefont{Eremin}},
  \bibinfo{author}{\bibfnamefont{A.}~\bibnamefont{Shorikov}}, \bibnamefont{and}
  \bibinfo{author}{\bibfnamefont{V.~I.} \bibnamefont{Anisimov}},
  \bibinfo{journal}{JETP Lett.} \textbf{\bibinfo{volume}{84}},
  \bibinfo{pages}{650} (\bibinfo{year}{2006}).

\bibitem[{\citenamefont{Motohashi et~al.}(2003)\citenamefont{Motohashi, Ueda,
  Naujalis, Tojo, Terasaki, Atake, Karppinen, and Yamauchi}}]{mot03}
\bibinfo{author}{\bibfnamefont{T.}~\bibnamefont{Motohashi}},
  \bibinfo{author}{\bibfnamefont{R.}~\bibnamefont{Ueda}},
  \bibinfo{author}{\bibfnamefont{E.}~\bibnamefont{Naujalis}},
  \bibinfo{author}{\bibfnamefont{T.}~\bibnamefont{Tojo}},
  \bibinfo{author}{\bibfnamefont{I.}~\bibnamefont{Terasaki}},
  \bibinfo{author}{\bibfnamefont{T.}~\bibnamefont{Atake}},
  \bibinfo{author}{\bibfnamefont{M.}~\bibnamefont{Karppinen}},
  \bibnamefont{and} \bibinfo{author}{\bibfnamefont{H.}~\bibnamefont{Yamauchi}},
  \bibinfo{journal}{Phys.\ Rev.\ B} \textbf{\bibinfo{volume}{67}},
  \bibinfo{pages}{064406} (\bibinfo{year}{2003}).

\bibitem[{\citenamefont{Schulze et~al.}(2008)\citenamefont{Schulze, H\"afliger,
  Niedermayer, Mattenberger, Bubenhofer, and Batlogg}}]{schu07}
\bibinfo{author}{\bibfnamefont{T.~F.} \bibnamefont{Schulze}},
  \bibinfo{author}{\bibfnamefont{P.~S.} \bibnamefont{H\"afliger}},
  \bibinfo{author}{\bibfnamefont{C.}~\bibnamefont{Niedermayer}},
  \bibinfo{author}{\bibfnamefont{K.}~\bibnamefont{Mattenberger}},
  \bibinfo{author}{\bibfnamefont{S.}~\bibnamefont{Bubenhofer}},
  \bibnamefont{and} \bibinfo{author}{\bibfnamefont{B.}~\bibnamefont{Batlogg}},
  \bibinfo{journal}{Phys.\ Rev.\ Lett.} \textbf{\bibinfo{volume}{100}},
  \bibinfo{pages}{026407} (\bibinfo{year}{2008}).

\bibitem[{\citenamefont{Yokoi et~al.}(2005)\citenamefont{Yokoi, Moyoshi,
  Kobayashi, Soda, Yasui, Sato, and Kakurai}}]{yok05}
\bibinfo{author}{\bibfnamefont{M.}~\bibnamefont{Yokoi}},
  \bibinfo{author}{\bibfnamefont{T.}~\bibnamefont{Moyoshi}},
  \bibinfo{author}{\bibfnamefont{Y.}~\bibnamefont{Kobayashi}},
  \bibinfo{author}{\bibfnamefont{M.}~\bibnamefont{Soda}},
  \bibinfo{author}{\bibfnamefont{Y.}~\bibnamefont{Yasui}},
  \bibinfo{author}{\bibfnamefont{M.}~\bibnamefont{Sato}}, \bibnamefont{and}
  \bibinfo{author}{\bibfnamefont{K.}~\bibnamefont{Kakurai}},
  \bibinfo{journal}{J.\ Phys.\ Soc.\ Jpn} \textbf{\bibinfo{volume}{74}},
  \bibinfo{pages}{3046} (\bibinfo{year}{2005}).

\bibitem[{\citenamefont{Yoshizumi et~al.}(2007)\citenamefont{Yoshizumi,
  Okamoto, Muraoka, Kiuchi, Ichihara, Yamaura, and Hiroi}}]{yos07}
\bibinfo{author}{\bibfnamefont{D.}~\bibnamefont{Yoshizumi}},
  \bibinfo{author}{\bibfnamefont{Y.}~\bibnamefont{Okamoto}},
  \bibinfo{author}{\bibfnamefont{Y.}~\bibnamefont{Muraoka}},
  \bibinfo{author}{\bibfnamefont{Y.}~\bibnamefont{Kiuchi}},
  \bibinfo{author}{\bibfnamefont{M.}~\bibnamefont{Ichihara}},
  \bibinfo{author}{\bibfnamefont{J.}~\bibnamefont{Yamaura}}, \bibnamefont{and}
  \bibinfo{author}{\bibfnamefont{Z.}~\bibnamefont{Hiroi}},
  \bibinfo{journal}{J.~Phys.~Soc.~Jpn.} \textbf{\bibinfo{volume}{76}},
  \bibinfo{pages}{063705} (\bibinfo{year}{2007}).

\bibitem[{\citenamefont{Hasan et~al.}(2004)\citenamefont{Hasan, Chuang, Qian,
  Li, Kong, Kuprin, Fedorov, Kimmerling, Rotenberg, Rossnagel et~al.}}]{has04}
\bibinfo{author}{\bibfnamefont{M.~Z.} \bibnamefont{Hasan}},
  \bibinfo{author}{\bibfnamefont{Y.~D.} \bibnamefont{Chuang}},
  \bibinfo{author}{\bibfnamefont{D.}~\bibnamefont{Qian}},
  \bibinfo{author}{\bibfnamefont{Y.~W.} \bibnamefont{Li}},
  \bibinfo{author}{\bibfnamefont{Y.}~\bibnamefont{Kong}},
  \bibinfo{author}{\bibfnamefont{A.}~\bibnamefont{Kuprin}},
  \bibinfo{author}{\bibfnamefont{A.~V.} \bibnamefont{Fedorov}},
  \bibinfo{author}{\bibfnamefont{R.}~\bibnamefont{Kimmerling}},
  \bibinfo{author}{\bibfnamefont{E.}~\bibnamefont{Rotenberg}},
  \bibinfo{author}{\bibfnamefont{K.}~\bibnamefont{Rossnagel}},
  \bibnamefont{et~al.}, \bibinfo{journal}{Phys.\ Rev.\ Lett.}
  \textbf{\bibinfo{volume}{92}}, \bibinfo{pages}{246402}
  (\bibinfo{year}{2004}).

\bibitem[{\citenamefont{Qian et~al.}(2006{\natexlab{b}})\citenamefont{Qian,
  Hsieh, Wray, Chuang, Fedorov, Wu, Luo, Wang, Viciu, Cava et~al.}}]{qui06b}
\bibinfo{author}{\bibfnamefont{D.}~\bibnamefont{Qian}},
  \bibinfo{author}{\bibfnamefont{D.}~\bibnamefont{Hsieh}},
  \bibinfo{author}{\bibfnamefont{L.}~\bibnamefont{Wray}},
  \bibinfo{author}{\bibfnamefont{Y.~D.} \bibnamefont{Chuang}},
  \bibinfo{author}{\bibfnamefont{A.}~\bibnamefont{Fedorov}},
  \bibinfo{author}{\bibfnamefont{D.}~\bibnamefont{Wu}},
  \bibinfo{author}{\bibfnamefont{J.~L.} \bibnamefont{Luo}},
  \bibinfo{author}{\bibfnamefont{N.~L.} \bibnamefont{Wang}},
  \bibinfo{author}{\bibfnamefont{L.}~\bibnamefont{Viciu}},
  \bibinfo{author}{\bibfnamefont{R.~J.} \bibnamefont{Cava}},
  \bibnamefont{et~al.}, \bibinfo{journal}{Phys.\ Rev.\ Lett.}
  \textbf{\bibinfo{volume}{96}}, \bibinfo{pages}{216405}
  (\bibinfo{year}{2006}{\natexlab{b}}).

\bibitem[{\citenamefont{Singh}(2000)}]{sin00}
\bibinfo{author}{\bibfnamefont{D.~J.} \bibnamefont{Singh}},
  \bibinfo{journal}{Phys.\ Rev.\ B} \textbf{\bibinfo{volume}{61}},
  \bibinfo{pages}{13397} (\bibinfo{year}{2000}).

\bibitem[{\citenamefont{Huang et~al.}(2004{\natexlab{a}})\citenamefont{Huang,
  Foo, Pascal, Lynn, Toby, He, Zandbergen, and Cava}}]{hua04}
\bibinfo{author}{\bibfnamefont{Q.}~\bibnamefont{Huang}},
  \bibinfo{author}{\bibfnamefont{M.~L.} \bibnamefont{Foo}},
  \bibinfo{author}{\bibfnamefont{R.~A.} \bibnamefont{Pascal}},
  \bibinfo{author}{\bibfnamefont{J.~W.} \bibnamefont{Lynn}},
  \bibinfo{author}{\bibfnamefont{B.~H.} \bibnamefont{Toby}},
  \bibinfo{author}{\bibfnamefont{T.}~\bibnamefont{He}},
  \bibinfo{author}{\bibfnamefont{H.~W.} \bibnamefont{Zandbergen}},
  \bibnamefont{and} \bibinfo{author}{\bibfnamefont{R.~J.} \bibnamefont{Cava}},
  \bibinfo{journal}{Physical Review B} \textbf{\bibinfo{volume}{70}},
  \bibinfo{pages}{184110} (\bibinfo{year}{2004}{\natexlab{a}}).

\bibitem[{\citenamefont{Huang et~al.}(2004{\natexlab{b}})\citenamefont{Huang,
  Khaykovich, Chou, Cho, Lynn, and Lee}}]{hua04b}
\bibinfo{author}{\bibfnamefont{Q.}~\bibnamefont{Huang}},
  \bibinfo{author}{\bibfnamefont{B.}~\bibnamefont{Khaykovich}},
  \bibinfo{author}{\bibfnamefont{F.~C.} \bibnamefont{Chou}},
  \bibinfo{author}{\bibfnamefont{J.~H.} \bibnamefont{Cho}},
  \bibinfo{author}{\bibfnamefont{J.~W.} \bibnamefont{Lynn}}, \bibnamefont{and}
  \bibinfo{author}{\bibfnamefont{Y.~S.} \bibnamefont{Lee}},
  \bibinfo{journal}{Phys.\ Rev.\ B} \textbf{\bibinfo{volume}{70}},
  \bibinfo{pages}{134115} (\bibinfo{year}{2004}{\natexlab{b}}).

\bibitem[{\citenamefont{Huang et~al.}(2005)\citenamefont{Huang, Lynn, Toby,
  Foo, and Cava}}]{hua05}
\bibinfo{author}{\bibfnamefont{Q.}~\bibnamefont{Huang}},
  \bibinfo{author}{\bibfnamefont{J.~W.} \bibnamefont{Lynn}},
  \bibinfo{author}{\bibfnamefont{B.~H.} \bibnamefont{Toby}},
  \bibinfo{author}{\bibfnamefont{M.~L.} \bibnamefont{Foo}}, \bibnamefont{and}
  \bibinfo{author}{\bibfnamefont{R.~J.} \bibnamefont{Cava}},
  \bibinfo{journal}{J.\ Phys-Condens.\ Mat.} \textbf{\bibinfo{volume}{17}},
  \bibinfo{pages}{1831} (\bibinfo{year}{2005}).

\bibitem[{\citenamefont{de~Vaulx et~al.}(2005)\citenamefont{de~Vaulx, Julien,
  Berthier, Horvatic, Bordet, Simonet, Chen, and Lin}}]{devau05}
\bibinfo{author}{\bibfnamefont{C.}~\bibnamefont{de~Vaulx}},
  \bibinfo{author}{\bibfnamefont{M.~H.} \bibnamefont{Julien}},
  \bibinfo{author}{\bibfnamefont{C.}~\bibnamefont{Berthier}},
  \bibinfo{author}{\bibfnamefont{M.}~\bibnamefont{Horvatic}},
  \bibinfo{author}{\bibfnamefont{P.}~\bibnamefont{Bordet}},
  \bibinfo{author}{\bibfnamefont{V.}~\bibnamefont{Simonet}},
  \bibinfo{author}{\bibfnamefont{D.~P.} \bibnamefont{Chen}}, \bibnamefont{and}
  \bibinfo{author}{\bibfnamefont{C.~T.} \bibnamefont{Lin}},
  \bibinfo{journal}{Phys.\ Rev.\ Lett.} \textbf{\bibinfo{volume}{95}},
  \bibinfo{pages}{186405} (\bibinfo{year}{2005}).

\bibitem[{\citenamefont{Zhou et~al.}(2005)\citenamefont{Zhou, Gao, Ding, Lee,
  and Wang}}]{zhou05}
\bibinfo{author}{\bibfnamefont{S.}~\bibnamefont{Zhou}},
  \bibinfo{author}{\bibfnamefont{M.}~\bibnamefont{Gao}},
  \bibinfo{author}{\bibfnamefont{H.}~\bibnamefont{Ding}},
  \bibinfo{author}{\bibfnamefont{P.~A.} \bibnamefont{Lee}}, \bibnamefont{and}
  \bibinfo{author}{\bibfnamefont{Z.~Q.} \bibnamefont{Wang}},
  \bibinfo{journal}{Phys.\ Rev.\ Lett.} \textbf{\bibinfo{volume}{94}},
  \bibinfo{pages}{206401} (\bibinfo{year}{2005}).

\bibitem[{\citenamefont{Laverock et~al.}(2007)\citenamefont{Laverock, Dugdale,
  Duffy, Wooldridge, Balakrishnan, Lees, Zheng, Chen, Lin, Andrejczuk
  et~al.}}]{lav07}
\bibinfo{author}{\bibfnamefont{J.}~\bibnamefont{Laverock}},
  \bibinfo{author}{\bibfnamefont{S.~B.} \bibnamefont{Dugdale}},
  \bibinfo{author}{\bibfnamefont{J.~A.} \bibnamefont{Duffy}},
  \bibinfo{author}{\bibfnamefont{J.}~\bibnamefont{Wooldridge}},
  \bibinfo{author}{\bibfnamefont{G.}~\bibnamefont{Balakrishnan}},
  \bibinfo{author}{\bibfnamefont{M.~R.} \bibnamefont{Lees}},
  \bibinfo{author}{\bibfnamefont{G.~Q.} \bibnamefont{Zheng}},
  \bibinfo{author}{\bibfnamefont{D.}~\bibnamefont{Chen}},
  \bibinfo{author}{\bibfnamefont{C.~T.} \bibnamefont{Lin}},
  \bibinfo{author}{\bibfnamefont{A.}~\bibnamefont{Andrejczuk}},
  \bibnamefont{et~al.}, \bibinfo{journal}{Phys.\ Rev.\ B}
  \textbf{\bibinfo{volume}{76}}, \bibinfo{pages}{052509}
  (\bibinfo{year}{2007}).

\bibitem[{\citenamefont{Ishida et~al.}(2005)\citenamefont{Ishida, Johannes, and
  Liebsch}}]{ish05}
\bibinfo{author}{\bibfnamefont{H.}~\bibnamefont{Ishida}},
  \bibinfo{author}{\bibfnamefont{M.~D.} \bibnamefont{Johannes}},
  \bibnamefont{and} \bibinfo{author}{\bibfnamefont{A.}~\bibnamefont{Liebsch}},
  \bibinfo{journal}{Phys.\ Rev.\ Lett.} \textbf{\bibinfo{volume}{94}},
  \bibinfo{pages}{196401} (\bibinfo{year}{2005}).

\bibitem[{\citenamefont{Roger et~al.}(2007)\citenamefont{Roger, Morris,
  Tennant, Gutmann, Goff, Hoffmann, Feyerherm, Dudzik, Prabhakaran, Boothroyd
  et~al.}}]{rog07}
\bibinfo{author}{\bibfnamefont{M.}~\bibnamefont{Roger}},
  \bibinfo{author}{\bibfnamefont{D.~J.~P.} \bibnamefont{Morris}},
  \bibinfo{author}{\bibfnamefont{D.~A.} \bibnamefont{Tennant}},
  \bibinfo{author}{\bibfnamefont{M.~J.} \bibnamefont{Gutmann}},
  \bibinfo{author}{\bibfnamefont{J.~P.} \bibnamefont{Goff}},
  \bibinfo{author}{\bibfnamefont{J.~U.} \bibnamefont{Hoffmann}},
  \bibinfo{author}{\bibfnamefont{R.}~\bibnamefont{Feyerherm}},
  \bibinfo{author}{\bibfnamefont{E.}~\bibnamefont{Dudzik}},
  \bibinfo{author}{\bibfnamefont{D.}~\bibnamefont{Prabhakaran}},
  \bibinfo{author}{\bibfnamefont{A.~T.} \bibnamefont{Boothroyd}},
  \bibnamefont{et~al.}, \bibinfo{journal}{Nature}
  \textbf{\bibinfo{volume}{445}}, \bibinfo{pages}{631} (\bibinfo{year}{2007}).

\bibitem[{\citenamefont{Zandbergen et~al.}(2004)\citenamefont{Zandbergen, Foo,
  Xu, Kumar, and Cava}}]{zand04}
\bibinfo{author}{\bibfnamefont{H.~W.} \bibnamefont{Zandbergen}},
  \bibinfo{author}{\bibfnamefont{M.~L.}~\bibnamefont{Foo}},
  \bibinfo{author}{\bibfnamefont{Q.}~\bibnamefont{Xu}},
  \bibinfo{author}{\bibfnamefont{V.}~\bibnamefont{Kumar}}, \bibnamefont{and}
  \bibinfo{author}{\bibfnamefont{R.~J.} \bibnamefont{Cava}},
  \bibinfo{journal}{Phys.\ Rev.\ B} \textbf{\bibinfo{volume}{70}},
  \bibinfo{pages}{024101} (\bibinfo{year}{2004}).

\end{thebibliography}

\end{document}